\documentclass{elsart1p}

 \usepackage{epsfig}

\usepackage{amssymb}

\def\bg{\begin{eqnarray}}
\def\nd{\end{eqnarray}}

\begin{document}

\begin{frontmatter}



\title{The Double Life of Thermal QCD}


\author{Mohammed Mia, Keshav Dasgupta, Charles Gale, and Sangyong Jeon}

\address{Department of Physics, McGill University, 3600 rue University, 
Montr\'eal, QC, Canada H3A 2T8}

\begin{abstract}
We study the gravity dual of a thermal gauge theory whose behavior parallels that of thermal QCD in the far IR.
The UV of our theory has infinite degrees of freedom. We
holographically renormalize the supergravity action to 
compute the stress tensor of the
dual gauge theory with fundamental flavors
incorporating the logarithmic running of the gauge
coupling. From the stress tensor we obtain the shear viscosity and the entropy of the medium at a temperature T, 
and investigate the violation of the bound for the viscosity to the entropy ratio. 
This is a shortened and simplified companion paper to hep-th/0902.1540, and is based on the talks given by 
M. Mia at the ``Strong and Electroweak Matter 2008'' workshop and the McGill workshop on ``AdS/CFT, Condensed 
Matter and QCD'' in the fall of 2008. 
\end{abstract}
\end{frontmatter}
\section{Introduction}
Strongly coupled Quark Gluon Plasma (QGP) poses theoretically challenging yet experimentally accessible questions. The
formation of QGP at RHIC is an example where theoretical descriptions are completely lacking at low energies 
because our   
perturbative techniques fail at strong couplings.  However
probing the nonperturbative  regime of large $N$ 
gauge theory through a {\it gravity} dual has led to
some interesting results for the physics of the quark gluon plasma. 
A popular approach so far has been the AdS/CFT correspondence \cite{maldacena}, 
even though QCD is not a conformal field theory at UV. 
However, for certain gauge theories with running couplings, 
there exist gravity duals. At zero temperatures these gravity duals are studied with \cite{ouyang}
and without \cite{strassler, strasslerreview} fundamental flavors. On the other hand,
at high temperatures, there are examples of gravity duals without fundamental flavors in the literature \cite{ahabu}. 
In this paper 
we give an example of a gravity dual for a specific 
thermal gauge theory with fundamental flavors and logarithmically running coupling constants (see also \cite{cotrone}). 
Our aim here is to study quantities such as shear viscosity, entropy and the viscosity to entropy 
ratio with our gravity dual.    
\section{The Duality}
\label{bla2}
The gauge theory that
we consider arises from stacking $N$ D3 branes at the tip of a six dimensional conifold with base 
$T^{1,1}=S^2\times S^3$, then
placing $M$ D5 branes in such a way that it wraps the two cycle $S^2$ on the base of the conifold alongwith 
$N_f$ 
$D7$ branes embedded via so-called Ouyang embedding \cite{ouyang}
to introduce fundamental matter \cite{FEP}. Finally by euclideanising and periodically identifying the time coordinate
we can introduce temperature in the gauge theory. However the high temperature, i.e temperature above the 
deconfinement temperature that we would be interested in, will be generated by inserting a black hole in the 
dual gravitational backgound \cite{FEP}. 
The D3 brane world volume and the unwrapped directions of D5 and D7 branes  
extends in the four Minkowski directions. Strings can end on any of these branes and the excitation
of the strings are described by an $SU(N+M)\times SU(N)$ gauge group with $N_f$ flavors. If $g_1,g_2$ denote the gauge
couplings of $SU(N+M)$ and $SU(N)$, then they have nontrivial beta functions (at zero temperature)
$\frac{\partial}{\partial{\rm log}~\Lambda}\left[\frac{4\pi^2}{g_1^2} + \frac{4\pi^2}{g_2^2}\right] = 
- \frac{3N_f}{4}$ and 
$\frac{\partial}{\partial{\rm log}~\Lambda}\left[\frac{4\pi^2}{g_1^2} - \frac{4\pi^2}{g_2^2}\right] = 
3M\left(1 + \frac{3g_s N_f}{2\pi} ~{\rm log}\Lambda\right)$.
We see that the two gauge couplings run in opposite directions and in the IR $SU(N+M)$ flows to strong
coupling. Performing a Seiberg duality
transformation, we identify the strongly coupled $SU(N+M)$ with a weakly coupled $SU(N-(M-N_f))$ at the IR. 
We see that not only the number of colors are reduced, but the difference of the size of the gauge group now decreases 
from $M$ in \cite{strassler} 
to $M-N_f$. This difference will decrease by the increments of $N_f$ until it is smaller than 
or equal to $N_f$. Then there are two possible end points: (a) if $N$ is still greater than zero then we will have an 
approximately conformal theory, or (b) if $N$ decreases to zero but with finite $M$ left over then we will have a 
$SU(M)$ theory with $N_f$ flavors that confines in the far IR (see 
\cite{ouyang} for more details). The latter theory, or
more particularly the high temperature limit of the latter theory, is what we are interested in and henceforth we will 
only consider that\footnote{For a review of Seiberg dualities and cascading theories, 
see \cite{strasslerreview}. For a review on brane constructions for cascading theories see \cite{DM}.}. 

At zero temperature the dual description (assuming of course that 
in this limit the gauge theory decouples from gravity) involves D7 branes, deformed conifold and fluxes \cite{ouyang}. Due
to the underlying RG flows the supergravity description {\it does not} capture the 
choppy cascading nature of the gauge theories!
Rather the supergravity description only captures the smooth RG flows in the theory. This would also mean that the actual 
{\it number} of colors in the theory is rather 
subtle to define. These details are explored in \cite{strasslerreview, FEP}. 

Once we switch on a temperature in the gauge theory, the dual gravity description loses its simplicity. 
We can no longer claim that the fluxes, 
warp factor etc would remain unchanged. Even the internal manifold cannot remain a simple 
warped deformed conifold any more. 
All the internal spheres would get squashed, and at $r = 0$ there could be both resolution as well as deformation
of the two and three cycles respectively. Due to this complicated nature of our background 
our entire analysis in \cite{FEP} is based on the following limit:
\begin{eqnarray}\label{prelim} 
\left (g_s, ~g_s N_f, ~g^2_s M N_f, ~{g_s M^2\over N}\right) ~ \to ~ 0, ~~~~~
(g_s N, ~g_s M)~ \to ~ \infty
\end{eqnarray}
In \cite{FEP} we presented our results to ${\cal O}(g_sN_f, g_sM^2/N)$, 
and discuss how to extend this to higher orders in $g_sN_f$ and  $g_sM^2/N$. 
In the limit where the deformation parameter is small, we show that to ${\cal O}(g_sN_f, g_sM^2/N)$ we can 
analytically derive the background taking a resolved conifold geometry. The resolution parameter $a$
depends on $g_sN_f, g_sM^2/N$ as well as on the horizon radius $r_h$ i.e $a = a_0 + {\cal O}(g_sN_f, g_sM^2/N, r_h)$ with
$a_0$ constant. 
Our background then has the following form:
\bg\label{bhmet1}
ds^2 &=& {1\over \sqrt{h}}
\Big[-g_1(r)dt^2+dx^2+dy^2+dz^2\Big] +\sqrt{h}\Big[g_2(r)^{-1}dr^2+ d{\cal M}_5^2\Big]\nonumber\\
{\widetilde F}_3 &=& 2M {\bf A_1} \left(1 + {3g_sN_f\over 2\pi}~{\rm log}~r\right) ~e_\psi \wedge 
\frac{1}{2}\left({\rm sin}~\theta_1~ d\theta_1 \wedge d\phi_1-{\bf B_1}~{\rm sin}~\theta_2~ d\theta_2 \wedge
d\phi_2\right)\nonumber\\
&& -{3g_s MN_f\over 4\pi} {\bf A_2}~{dr\over r}\wedge e_\psi \wedge \left({\rm cot}~{\theta_2 \over 2}~{\rm sin}~\theta_2 ~d\phi_2 
- {\bf B_2}~ {\rm cot}~{\theta_1 \over 2}~{\rm sin}~\theta_1 ~d\phi_1\right)\nonumber\\
&& -{3g_s MN_f\over 8\pi}{\bf A_3} ~{\rm sin}~\theta_1 ~{\rm sin}~\theta_2 \left({\rm cot}~{\theta_2 \over 2}~d\theta_1 +
{\bf B_3}~ {\rm cot}~{\theta_1 \over 2}~d\theta_2\right)\wedge d\phi_1 \wedge d\phi_2\nonumber\\
H_3 &=&  {6g_s {\bf A_4} M}\Bigg(1+\frac{9g_s N_f}{4\pi}~{\rm log}~r+\frac{g_s N_f}{2\pi} 
~{\rm log}~{\rm sin}\frac{\theta_1}{2}~
{\rm sin}\frac{\theta_2}{2}\Bigg)\frac{dr}{r}\wedge \frac{1}{2}\Big({\rm sin}~\theta_1~ d\theta_1 \wedge d\phi_1\nonumber\\
&&-{\bf B_4}~{\rm sin}~\theta_2~ d\theta_2 \wedge d\phi_2\Big)
+ \frac{3g^2_s M N_f}{8\pi} {\bf A_5} {\cal D}(r, \psi) 
\wedge \Bigg({\rm cot}~\frac{\theta_2}{2}~d\theta_2 
-{\bf B_5}~{\rm cot}~\frac{\theta_1}{2} ~d\theta_1\Bigg)\nonumber\\
C_0 & = & {N_f \over 4\pi} (\psi - \phi_1 - \phi_2), ~~~~~~
F_5 ~ = ~ {1\over g_s} \left[ d^4 x \wedge d h^{-1} + \ast(d^4 x \wedge dh^{-1})\right]\nonumber\\
e^{-\Phi} & = & {1\over 2g_s}\left[{1\over r^{\epsilon_a}} - {3\epsilon_a a^2\over 2 r^2} 
-\frac{g_sN_f}{2\pi} {\rm log} \left({\rm sin}~{\theta_1\over 2} ~ {\rm sin}~{\theta_2\over 2}\right) 
+ {\rm constant}\right]
\nd  
where ${\cal D}(r, \psi) \equiv \frac{dr}{r}\wedge e_\psi -\frac{1}{2}de_\psi$; $g_i$ are the black 
hole factors\footnote{Note that they are generically unequal.};
$\epsilon_a = {3g_sN_f \over 4\pi}$; 
${\bf A}_i, {\bf B}_i$ are 
${\cal O}(a^2, g_sN_f)$ corrections that take us away from the zero temperature
Ouyang background \cite{ouyang}, and are worked out in details in \cite{FEP}; and  
${\cal M}_5$ is a warped {\it resolved}
conifold with the warp factor:
\bg\label{logr}
h~=~ {L^4\over r^{4- \epsilon_1}} ~+~ {L^4\over r^{4- 2\epsilon_2}} ~-~ {2L^4\over r^{4- \epsilon_2}} ~+~ {L^4 \over 
r^{4-r^{\epsilon^2_2/2}}} 
~\equiv~ \sum_{\alpha=1}^4 {L^4_{(\alpha)}\over r^4_{(\alpha)}}
\nd
where $\epsilon_i, r_{(\alpha)}$ etc are defined as:
\bg\label{epde}
&&\epsilon_1 ~ = ~ {3g_s M^2\over 2\pi N} + {g_s^2 M^2 N_f\over 8\pi^2 N} + {3g_s^2 M^2 N_f \over 8\pi N} ~
{\rm log}\left({\rm sin}~{\theta_1\over 2} {\rm sin}~{\theta_2\over 2}\right),  
~~\epsilon_2 ~ = ~ {g_s M \over \pi}\sqrt{2N_f\over N}\nonumber\\
&& ~~~~~~~~~~~ r_{(\alpha)} = r^{1-\epsilon_{(\alpha)}}, ~~ \epsilon_{(1)} = {\epsilon_1\over 4}, ~~ 
\epsilon_{(2)} = {\epsilon_2\over 2}, ~~ \epsilon_{(3)} = {\epsilon_2\over 4}, ~~ \epsilon_{(4)} = 
{\epsilon^2_2\over 8}\nonumber\\
&& ~~~~~~~~~~~ r_{(\pm\alpha)} = r^{1\mp\epsilon_{(\alpha)}}, ~~~L_{(1)} = L_{(2)} = L_{(4)} = L^4, ~~~ L_{(3)} = -2L^4 
\nd
and $L^4$ is defined 
in the same way as in the AdS/CFT correspondence. The D7 branes are embedded via Ouyang embedding in this background. 
Note that our analysis to this order is analytic, but 
beyond this order we loose all analytic control and we can only find the 
background numerically (see also \cite{ahabu} for the background without flavor D7 branes and \cite{haack} for the 
background with flavor branes). 
Furthermore the 
${\rm log}~r$ dependences of all the fluxes, warp factor and dilaton is only for regions close to the D7 brane. To get 
a good behavior at far infinity we need to embed the whole system in F-theory \cite{vafaF}. That would mean 
inserting extra $24 - N_f$ seven branes at infinity; and then all the fluxes would go as inverse powers of $r$ at 
large $r$. In fact in \cite{FEP} we made a concrete prediction for the large $r$ warp factor: it is given by 
(\ref{logr}) except that the sum over $\alpha$ now ranges from $-\infty$ to $+\infty$. 
This way we can get rid of possible Landau poles (or naked singularities) in the background (see also \cite{vaman}). 
On the other hand for 
{\it finite} $g_sN_f, g_sM^2/N$ the small $r$ behavior is almost identical to (\ref{logr}). However we 
couldn't derive the precise form for the large $r$ behavior of the warp factor. Therefore in \cite{FEP} we only 
speculated a possible inverse $r$ behavior for the warp factor.  

Once we know the dual background we can say that the
Hilbert space of the gauge theory can be obtained from the Hilbert space of the string theory on this 
geometry. In fact the dual supergravity background captures the strongly coupled gauge theory 
where we expect the RG flow to be smooth without any choppy Seiberg dualities, although we might be able to interpret
the strongly coupled gauge theory also as some kind of approximate cascading theory.

As we now know very well, 
the UV completion of cascading type theories require {\it infinite} 
degrees of freedom. Once we have infinite degrees of freedom at the UV, we no longer 
expect a finite boundary action from supergravity analysis! What we need is to regularise and renormalise the 
supergravity boundary action so that finite correlation functions could be extracted. This would also mean that 
the usual Witten type proposal \cite{witten} for the AdS/CFT correspondence can be re-expressed
in terms of the boundary variables to give us the complete picture.    
Therefore we can rewrite the ansatze proposed by Witten {\it et al} \cite{witten}
for our background geometry to take the following Wilsonian form:
\bg \label{KS16}
{\cal Z}_{\rm QCD}[\phi_0] &~\equiv &~ \langle{\rm exp}\int_{M^4} \phi_0 {\cal O}\rangle
~=~ {\cal Z}_{\rm total}[\phi_0] \nonumber\\
&~ \equiv &  ~{\rm exp}(S_{\rm total}[\phi_0] + S_{\rm GH} + S_{\rm counterterm}) ~\equiv~ {\rm exp}~S_{\rm ren}[\phi_0]
\nd
where $M^4$ is Minkowski manifold, $S_{\rm total}$ is the low energy type IIB action defined in the string 
frame\footnote{There are subtleties involved in using $S^{\rm total}_{\rm string}\simeq S^{\rm total}_{\rm SUGRA}$. 
For detailed discussions, consult \cite{FEP}.}, $\phi_0$ is the fluctuation over a given background, 
$S_{\rm GH}$ is the Gibbons-Hawking boundary term \cite{gibhaw}
and $S_{\rm counterterm}$ is the 
counter-term action added to renormalise the action. 

The interesting part now is that we can define {\it two} classes of theories in this kind of backgrounds by 
introducing a cut-off at $r = r_c$:

\noindent $\bullet$ The first class 
is to analyse the theory right at the 
usual boundary 
where $r_c \to \infty$. This is the standard picture where there are infinite degrees of freedom at the boundary, and 
the theory has a smooth RG flow from UV to IR till it confines (at least from the weakly coupled gravity dual). 

\noindent $\bullet$ The 
second class is to analyse theories by specifying the 
degrees of freedom at generic $r_c$ and then defining the theories at the boundary. All these theories would meet the 
cascading theory at certain scales under RG flows. The gravity duals of these theories are the usual 
resolved-deformed conifold 
geometries cut-off 
at various $r_c$ with appropriate UV caps added (or, alternatively, degrees of freedom specified). These
UV caps are non-trivial geometries added to the 
resolved-deformed conifold geometry from $r = r_c$ to $r = \infty$\footnote{For
example we could add AdS geometries from $r = r_c$ to $r = \infty$. This is like adding ${\cal N} = 4$ or 2 
degrees of freedom
at the cut-off scale to UV complete the gauge theories. One issue here is 
the connection to the work of \cite{balkrauss}. The question is 
can we write a boundary theory at $r = r_c$ itself instead of going to the 
actual $r = \infty$ boundary? As shown in \cite{balkrauss} this is possible in AdS case because the 
theory on the surface of a ball in AdS space doesn't have to be a local quantum field theory. The non-local behavior
in such a ``boundary'' theory is completely captured by the Wilsonian effective action at the so called boundary. 
As detailed in \cite{FEP},
this is tricky in the Klebanov-Strassler model precisely because there is no unique strongly 
coupled gauge theory here. Again, as in the RG flow pictures of \cite{FEP}, 
when the dual gravity theory is weakly coupled 
we have a smooth RG flow in the gauge theory side, but the theory at any given scale can be given by 
infinite number of representative gauge theories none of which completely capture the full dynamics. Thus it makes more 
sense to 
define the theories at $r = \infty$ boundary only and not at any generic $r = r_c$.}. In addition to that, both these
classes of theories ought to have appropriate number of seven branes so that all the Landau poles could be removed.  

We can also say that the boundary degrees of freedom are now given by ${\cal N}_{\rm uv}$ where 
${\cal N}_{\rm uv} = \epsilon^{-n}, \epsilon \to 0$ and $n >> 1$ 
specifies the parent cascading theory. For all other theories that we define at the 
boundary $r = \infty$, by 
cutting off the geometry at $r = r_c$, will have ${\cal N}_{\rm uv}$ going as ${\cal N}_{\rm uv} = \epsilon^{-n}$ but
now $n \ge 1$. 
Yet another interesting thing 
about all these theories is that {\it any} thermodynamical quantitities defined in these theories will be 
completely independent of the cut-off $r_c$ and only depend on the temperature $T$ (with an exponentially small 
dependences
on the boundary degrees of freedom)! This will be briefly demonstrated in the next section (see \cite{FEP} for the 
full exposition). 
\begin{figure}[htb]\label{cascade}
\begin{center}
\includegraphics[height=3.8cm,width=7cm]{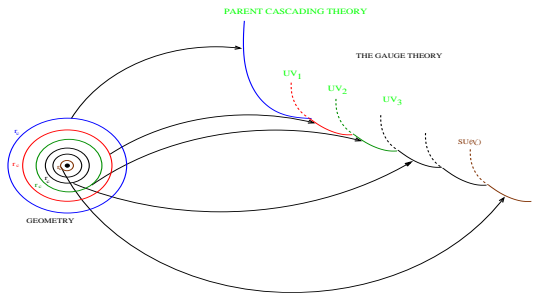}
\caption{{A very approximate way to denote all the UV completed theories that we 
could study in our framework by cutting off the geometry at various $r = r_c$. 
For a more correct illustration of the situation see {\cite{FEP}}.  
The outermost blue curves on both sides
denote the parent cascading theory.}}
\end{center}
\end{figure} 

\section{The Stress Tensor}
\label{bla3}
One of the important thermodynamical quantity to extract from our background would be the stress tensor.  
As the stress tensor couples to the metric, it's expectation value is obtained from the renormalised action $S_{\rm ren}$
via 
$\langle T^{ij}(\Lambda_c)\rangle=\frac{\delta S_{\rm ren}[l_{ij}]}{\delta l_{ij}}\Big{|}_{l_{ij}=0}$
where $l_{ij}$ is the four dimensional effective metric induced by the ten dimensional metric of the
 form (\ref{bhmet1}) by approximating the metric first by a five dimensional slice and then taking the required 
pull-back (see \cite{FEP} for details on the procedure). 
At every fixed $r$ the five dimensional metric induces a four dimensional metric and with appropriate rescaling of the 
time coordinate, we can view this effective
four dimensional metric as a flat Minkowski metric (with possible perturbations). 
If $O_{ij}$ denote the perturbations on the
background by quark strings that stretch from the D7 branes to the horizon of the black hole,
then writing the supergravity action up to quadratic order in $O_{ij}$; using integration by 
parts together with appropriate Gibbons-Hawking terms; and holographically renormalising the subsequent action, we obtain:
\bg\label{wakeup}
&& T^{mm}_{{\rm medium} + {\rm quark}}    
 = \int \frac{d^4q}{(2\pi)^4}\sum_{\alpha, \beta}
\Bigg\{({H}_{\vert\alpha\vert}^{mn}+ {H}_{\vert\alpha\vert}^{nm})s_{nn}^{(4)[\beta]} 
-4({K}_{\vert\alpha\vert}^{mn}+ {K}_{\vert\alpha\vert}^{nm})s_{nn}^{(4)[\beta]}\nonumber\\
&& ~~~~~~~~~~~ +({K}_{\vert\alpha\vert}^{mn}+ {K}_{\vert\alpha\vert}^{nm})s_{nn}^{(5)[\beta]}
+\sum_{j=0}^{\infty}~\hat{b}^{(\alpha)}_{n(j)} \widetilde{J}^n  
\delta_{nm}  e^{-j{\cal N}_{\rm uv}} + {\cal O}({\cal T} e^{-{\cal N}_{uv}})\Bigg\}
\nd 
where ($\hat{b}^{(\alpha)}_{n(j)}, 
{\cal N}_{\rm uv}$) together will specify the full boundary theory for a specific
UV complete theory.
The quantities $H^{mn},\widetilde{J}^{m},K^{mn}$ etc. with 
$m,n=0,..,3$ are completely independent of the radial coordinate $r$ and depend on the 3+1 dimensional spacetime 
coordinates. Note also that the stress tensor depends only on the temperature $T$ and is independent of any cut-offs. The 
dependence on the UV degrees of freedom is exponentially suppressed, and in the limit 
${\cal N}_{\rm uv} = \epsilon^{-n}, n >> 1$
we reproduce the result for the parent cascading theory.  

\section{Results and Discussion}
\label{bla3}
With the
general formulation of stress tensor (\ref{wakeup}), 
which is similar to the AdS/CFT results \cite{Skenderis} in the limiting case $M= N_f=0$,
we can compute the wake a moving quark creates in a plasma with $O_{ij}$ being the
metric perturbation due to string (see \cite{Yaffe} for an equivalent AdS calculation). 
Furthermore we can compute the shear viscosity $\eta$ from the Kubo formula
with the propagator obtained from the dual partition function (\ref{KS16}) and the entropy density
$s$ using Wald's formula. 
The result for the ratio $\eta/s$ is given by:
\bg \label{final1}
\frac{\eta}{s} &=&~ {1 + \sum_{k = 1}^\infty
\alpha_k e^{-4k {\cal N}_{\rm uv}} \over 4\pi + {1\over \pi}~{\rm log}^2 \left(1 - 
{T}^4 e^{-4{\cal N}_{uv}}\right)}\nonumber\\
&-&\frac{c_3\kappa}{3 L^2 \left(1- {T}^4 e^{-4{\cal N}_{uv}}\right)^{3/2}}
 \left[\frac{{B_o}(4\pi^2-{\rm log}^2 ~C_o)+4\pi{A_o}~{\rm log}~C_o}{\Big(4\pi^2-{\rm
log}^2~C_o\Big)^2+16\pi^2~{\rm
log}^2~C_o}\right]
\nd
where ($A_o, B_o, C_o, \alpha_k$) are constants that depend on the temperature $T$ and $e^{-{\cal N}_{\rm uv}}$; and 
$c_3$ is the coefficient of the Riemann square term coming from the backreactions of the D7 branes in the background.
These have been explicitly worked out in \cite{FEP}. The $c_3$ dependence of the $\eta/s$ ratio first appeared 
in \cite{kats}. 

With certain choices of parameters ($g_s = 0.01, N=1000, M=100, N_f=3, c_3=0.001$) and ignoring 
${\cal O}(g_sN_f, g_sM^2/N)$ corrections to the black hole parameters $g_i(r)$  
the result for $\eta/s$ is shown in {figure 2}. It is easy to see that we are violating the celebrated KSS 
bound \cite{Kovtun}\footnote{Recent papers dealing with the violation of the KSS bound are \cite{kats, violation}.}.  
\begin{figure}[htb]\label{ETAoS}
		\begin{center}
                \includegraphics[height=8cm,angle=-90]{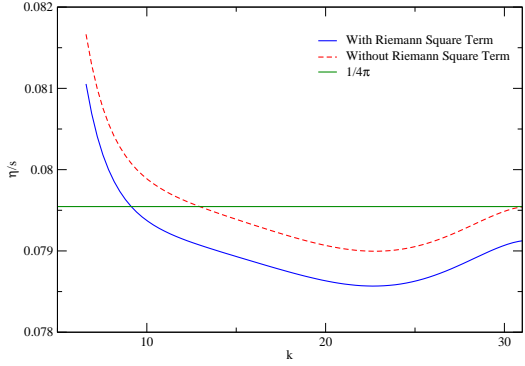}
\caption{Plot of $\eta/s$ versus the UV degrees of freedom.}		
		\end{center}
		\end{figure}
The plot in {figure 2} is shown for 
$\eta/s$ with and without Riemann Square term. The x-axis is defined as 
$k = {e^{{\cal N}_{\rm uv}}\over {T}}$ where ${\cal N}_{\rm uv}$ is the UV degrees of freedom and 
${T}$ is the temperature of the cascading theories. For the parent cascading 
theory, and their corresponding family of theories, $k \to \infty$
and we see a violation of the bound (the solid blue line). As $k$ decreases (assuming this is possible!) the red dashed 
line dips slightly below the $1/4\pi$ axis, but the solid blue line remains considerably below the $1/4\pi$ axis. For 
$k$ sufficiently small the bound is not violated. However all the models that we studied here and in 
\cite{FEP} can {\it only} realise the infinite $k$ limit, so that the small $k$ limit depicted in Fig. 2 above is just an 
extrapolation of (\ref{final1}).

\section{Conclusion}
We have analyzed the gravity dual of a thermal
gauge theory with logarithmic running coupling which may resemble QCD in the far IR.
It appears that the KSS bound for $\eta/s$ may be violated in certain gauge theories if one carefully takes into account
both the running of the couplings as well as the backreactions of the flavor branes in the dual gravity picture, for a range of parameter values (see details in \cite{FEP}).
It is of course important to test and verify the robustness of this limit, and we view the current work as contributing to this effort.  More work is needed in order to identify the size and extent of this violation, at the moment this violation is only parametric and our parameters need to have better defined physical origins. In the end, one may need to rely on the empirical identification of key quantities, like transport coefficients for example \cite{Luzum}. In this regard, the role of heavy ion experiments at RHIC and at the LHC can't be overestimated. 

\section{Acknowledgments}
We would like to thank Alex Buchel, Paul Chesler, Andrew Frey, Jaume Gomis, Evgeny Kats, Ingo Kirsch,  
Rob Myers, Peter Ouyang, Omid Saremi, Aninda Sinha, Diana Vaman, Larry Yaffe and especially 
Ofer Aharony and Matt Strassler for many helpful discussions and 
correspondence.
This work is 
supported in part by the Natural Sciences and Engineering Research 
Council (NSERC) of Canada, and in part by McGill University.



\end{document}